# Comparison of three methods to compute optical absorption spectra of organic molecules and solids

Ang Siong Tuan, Amrita Pal, Sergei Manzhos[1]

Department of Mechanical Engineering, National University of Singapore, Block EA #07-08, 9 Engineering Drive 1, Singapore 117576

**Abstract**

We compare the performance of an approach using real frequency dependent polarizability to compute optical absorption spectra to linear-response time-dependent density functional theory (TD-DFT) for small organic dyes, oligomers of different length (oligothiophenes), and molecular clusters representing a molecular crystal (pentacene). For pentacene, the spectra computed with the two methods are also compared to the spectrum computed for clusters and the periodic solid using the dipole approximation. The approach based on real polarizability produces spectra in good agreement with TD-DFT for small molecules. The (artificial) redshift for longer oligomers is more significant with the polarizability-based method than with TD-DFT. For pentacene clusters, TD-DFT produces reasonable spectra with a hybrid functional, but a significant redshift is introduced with a GGA functional due to the presence of charge transfer transitions. This problem is slightly attenuated with the polarizability-based method. The dipole approximation results in spectra much redshifted vs both TD-DFT and the polarizability-based method and in a different trend with cluster size.

**Introduction**

Absorption and emission spectra of organic materials in applications such as organic and dye-sensitized solar cells or organic light-emitting diodes are critically important material properties.[1-6] Experimental spectra for materials characterization are usually measured either in solution, in which case they are usually dominated by single-molecule properties, or in in solid or thin film state, in which case they may be strongly influenced by inter-molecular interactions. We note that in most applications, it is the spectrum in solid state that is practically relevant. Computing such spectra is important for rationalization of material and device performance as well as for prediction of material properties and computational materials design. The quality of computed spectra is therefore practically, including economically, important.

---

[1] Author to whom correspondence should be addressed. Tel: +65 6516 4605; fax: +65 6779 1459; E-mail: mpemanzh@nus.edu.sg

When computing optical absorption spectra for these applications, by far the most widely used approach is linear-response Time-Dependent Density Functional Theory[7, 8] (TD-DFT, we will imply linear response in what follows unless otherwise stated). Real-time TD-DFT has been developed but is not widely used in applied literature and is computationally costly. While TD-DFT is in principle applicable in solid state (periodic calculations), in practice, this is costly. Typically, one performs TD-DFT calculations on 0D systems, for which TD-DFT is implemented in widely used codes such as Gaussian.[9] Small nanoparticles are sometimes used to model solid state of inorganic materials[10, 11] and most often single-molecule calculations (molecules or oligomers) are used for organic materials.[12-14] Specifically for solid organic materials, on one hand, there may be significant effects on the spectrum due to aggregation, and on the other hand, such effects are more difficult to model with TD-DFT than single-molecule spectra, especially, when charge-transfer transitions between molecular units are introduced.[15-18] TD-DFT spectra of even single molecules are also in significant error when there is significant charge-transfer character or a significant extent of conjugation, in which case artificial and strong redshift is introduced.[19-21] Absolute errors in excitation energies need not be significant for the calculation to result in a significant and qualitative error; for example, in Ref. [22] we showed that a simple change in conjugation order which introduces significant changes in measured spectra is not qualitatively reproduced with TD-DFT using GGA (generalized gradient approximation) or simple hybrid functionals.[22-24] Such qualitative errors can be corrected by the use of range-separated hybrid functionals but at the price of losing the quantitative or semi-quantitative accuracy of the HOMO (highest occupied molecular orbital) and LUMO (lowest unoccupied molecular orbital) energy levels and of absorption peaks which is achieved by simple hybrid functionals such as B3LYP.[25, 26]

The errors in TD-DFT spectra relate to the way the spectrum is computed. The excitation spectrum $\omega$ is obtained from the generalized eigenvalue problem

$$\begin{bmatrix} A & B \\ B & A \end{bmatrix} \begin{bmatrix} X \\ Y \end{bmatrix} = \omega \begin{bmatrix} -1 & 0 \\ 0 & -1 \end{bmatrix} \begin{bmatrix} X \\ Y \end{bmatrix}, \qquad (1)$$

where the elements of matrices *A* and *B* depend on the integrals

$$K_{ia\mu,jb\nu} = \iint \psi_{i\sigma}^*(r)\psi_{a\sigma}(r)\left(\frac{1}{|r-r'|} + \frac{\delta^2 E_{XC}}{\delta\rho_\sigma(r)\delta\rho_\nu(r')}\right)\psi_{j\nu}(r')\psi_{b\nu}^*(r')drdr'$$

(2)

where indices *i, j* and *a, b* label occupied and virtual orbitals $\psi$, respectively, and indices $\sigma$ and $\nu$ denote spin, $\rho$ is the density, and $E_{XC}$ the exchange-correlation energy.[27] Eq. 2 is very sensitive to the quality of the orbitals and to any errors in the orbitals, in particular, because it involves overlap integrals with a kernel. As a result, the effect of errors in orbitals is much stronger on Eq. 2 than it is e.g. on orbital energies and depends significantly on functional choice.[22]

In periodic solid-state calculations, the so-called dipole approximation has been mostly used as a relatively inexpensive way to produce optical absorption spectra.[16, 28, 29] In it, the imaginary part of the frequency-dependent dielectric function $\epsilon_i(\omega)$ is computed using

$$\epsilon_i(\omega) = \frac{2e^2\pi}{\Omega\epsilon_0}\sum_{k,a,b}|\langle\psi_k^b|\hat{q}\cdot r|\psi_k^a\rangle|^2 \delta(E_k^b - E_k^a - \hbar\omega)$$

(3)

where $\Omega$ is the cell volume, indices *a* and *b* scan occupied and unoccupied $\psi_k^{a,b}$ orbitals (whose eigenstates are $E_k^{a,b}$), respectively, $k$ is the wavevector (in the Brillouin zone), and $q$ is the photon polarization vector. From the imaginary part of the frequency-dependent dielectric function, one can compute the real part using the Kramers-Kronig relation[30, 31]

$$\epsilon_r = -\frac{2}{\pi}\text{P}\int_0^\infty \frac{\omega'\epsilon_i(\omega')}{\omega^2 - \omega'^2}d\omega'$$

(4)

where P stands for the principal value.[32] The absorption spectrum (molar absorptivity $\mu$) is then computed as

$$M\mu(\omega) = \frac{\sqrt{2}\omega}{c}\left(\sqrt{\epsilon_r(\omega)^2 + \epsilon_i(\omega)^2} - \epsilon_r(\omega)\right)^{\frac{1}{2}}$$

(5)

where $M$ is molar concentration and $c$ the speed of light. The calculation of $\epsilon_i(\omega)$ and with it the absorption spectrum also critically relies on the shape of Kohn-Sham orbitals due to an overlap integral with a kernel. Eq. 3 as well as TD-DFT directly depend on the Kohn-Sham spectrum, i.e. the underestimation of the HOMO-LUMO gap with GGA functionals directly translates into a redshifted spectrum, while the use of hybrid functionals is not practical (too costly) in many applications using periodic calculations. Indeed, spectra of organic crystals computed with the dipole approximation exhibit significant unrealistic redshift.[33]

It is therefore desirable to explore alternative approaches to computing optical absorption spectra, which could have the potential to alleviate some of the drawbacks of TD-DFT and of the dipole approximation. In Ref. [33], when computing the spectrum of C60 and C60 clusters, we used an approach based on real polarizability $\alpha(\omega)$. The real part of the dielectric constant $\epsilon_r$ can be computed from $\alpha(\omega)$ using the Clausius–Mossotti relation[34, 35]

$$\frac{\epsilon_r - 1}{\epsilon_r + 2} = \frac{N\alpha}{3\epsilon_0}$$

(6)

where $N$ is the numbers density of molecules and $\epsilon_0$ the permittivity of vacuum. Then the imaginary part of the complex dielectric constant is computed using the Kramers-Kronig relation

$$\epsilon_i = \frac{2}{\pi} P \int_0^\infty \frac{\omega' \epsilon_r(\omega')}{\omega^2 - \omega'^2} d\omega'$$

(7)

and the absorption spectrum is computed with Eq. 5. The calculation of $\alpha(\omega)$, as it is implemented in the Gaussian code[9] used here, does depend on integrals over occupied and unoccupied Kohn-Sham orbitals $\langle \phi_a \phi_j | \phi_b \phi_i \rangle$, $\langle \phi_a \phi_b | \phi_i \phi_j \rangle$; however, the dependence can be less sensitive due to the absence of a kernel. It is also conceivable that orbital-independent approaches to compute $\alpha(\omega)$ will be implemented. This approach may therefore have potential to alleviate some of the errors introduced by the TD-DFT and the dipole approximations. It has other advantages and disadvantages; specifically, it is well parallelizable as $\alpha(\omega)$ can be computed for each frequency. On the other hand, it appears to be costlier than TD-DFT and suffers from instabilities when computing $\alpha(\omega)$ at high frequencies. So much so that in Ref. [33] we had to limit ourselves to a GGA functional and to dimers when computing the polarizability of C60 clusters.

In Ref. [33], we observed that this approach provides more realistic aggregation effects on the absorption spectrum of C60 than the dipole approximation and TD-DFT using a GGA functional. The absorption peaks of C60 are, contrary to most molecules used in solar cell and OLED applications, not dominated by transitions between frontier orbitals but are due to transitions involving many "deep" occupied and "deep" unoccupied orbitals. The performance of the polarizability-based method (Eqs. 5-7) remains unexplored for molecules and molecular aggregates in which transitions between frontier orbitals determine the visible absorption peak. In the present work, we explore that performance and compare it to TD-DFT for small dyes, oligomers, and molecular aggregates representing an organic crystal. For the latter case, we also compare to the periodic calculations using the dipole approximation. We use 2-cyano-3-[5'-(4''-(N,N-dimethylamino) phenyl) thiophen-2'-yl]-acrylic acid and cyano-[5-(4'-(N,N-dimethylamino) benzylidene)-5H-thiophen-2-ylidene]-acetic acid as examples of small dyes. In previous works,[22-24] it was shown that these two molecules show significant differences in the absorption peak maximum induced by the change in the conjugated order which cannot be accounted for qualitatively by GGA and simple hybrid functionals or quantitatively by a range-separated hybrid functional. We use thiophene oligomers of different lengths as examples of molecules with increasing degrees of conjugation where artificial redshift is expected. We use pentacene as an example of a molecular crystal, where effects of aggregation on the absorption spectrum are of interest (rather than single-molecule spectra in the previous two cases). We show that the polarizability based method shows performance similar to that of TD-DFT for small molecules. For oligomers, the redshift with degree of conjugation is stronger than with TD-DFT, i.e. the method underperforms. For pentacene clusters, there is slight advantage over TD-DFT, and significant advantage over the dipole approximation, when using a GGA functional.

**Methods**

DFT calculations on all molecules and clusters were performed in Gaussian 09[9] using the LANL2DZ basis set.[36] LANL2DZ provides a good balance of basis completeness, size and accuracy.[22, 33] PBE[37], B3LYP[25, 26], and ωB97xd[38] functionals were used as examples of widely used GGA, simple hybrid, and range-separated hybrid functionals. Not all functionals were used for each system. Default convergence criteria were used and were sufficient. TD-DFT calculations used 36, 36, and 50 states for small molecules, oligomers, and pentacene clusters, respectively. The states' excitation energies $E_i^{exc}$ and oscillator strengths $f_i$ obtained with TD-DFT were used to calculate the molar absorptivity $\mu$ as a continuous function of the excitation energy $E$ using

$$\mu(E) = \frac{1.35 \times 10^4}{\sigma} \sum_i f_i \, exp\left[-2.772\left(\frac{E - E_i^{exc}}{2\sigma}\right)^2\right]$$

(8)

with the HWHM (half width half maximum) broadening $\sigma = 0.25$ eV. The spectra computed from polarizability (Eqs. 5-7) as well as the spectra computed in periodic simulations were also broadened to a similar extent. To apply Eqs. 5-7, the polarizability was computed up to $\omega = 10$ eV with a step of 0.01 eV and then, for the purpose of the application of the Kramers-Kronig relation, padded with zeros up to 20 eV.

Periodic DFT calculations were performed in SIESTA[39] using the PBE functional.[37] A double-ζ polarized (DZP) basis set was used generated using the option PAO.EnergyShift = 0.002 Ry. The density was expanded using a plane wave cutoff of 150 Ry. The density matrix convergence criterion was $1 \times 10^{-5}$, the force convergence criterion was 0.02 eV/Å, and the pressure convergence criterion in solid state calculations was 0.01 GPa. Grimme dispersion corrections were used.[40] The unit cell was used to compute the crystalline pentacene,[41] and the Brillouin zone was sampled by 4×3×2 Γ-centered Monkhorst-Pack $k$-points.[42] The dipole approximation calculation of $\epsilon_i(\omega)$ (Eq. 3) is done in the "polycrystal" regime effectively averaging over $\boldsymbol{q}$. Monomer and dimer calculations were performed at the Γ point by placing the molecules in the center of a cubic simulation cell of size 30 Å, which was sufficient to neglect inter-cell interactions.

Calculations on pentacene clusters in Gaussian 09[9] and SIESTA[39] were performed by positioning optimized monomers with respect to each other as in the pentacene crystal.[41] All calculations are performed in vacuum.

**Results**

*Small dyes*

The structures of 2-cyano-3-[5'-(4''-(N,N-dimethylamino) phenyl) thiophen-2'-yl]-acrylic acid (which we will call Dye 1 in the following) and cyano-[5-(4'-(N,N-dimethylamino) benzylidene)-5H-thiophen-2-ylidene]-acetic acid (which we will call Dye 2) are shown in **FIG. 1**. Their absorption spectra computed with different methods are shown in **FIG. 2**, and the corresponding absorption peak maxima are listed in **Table 1**. These results suggest that the polarizability-based method provides a similar accuracy of the excitation energies of small molecules as TD-DFT, with differences between the two methods on the order of 0.1 eV within what is considered to be good accuracy for TD-DFT calculations.[17, 43] The dependence on the exchange-correlation functional is also similar in both methods; specifically, the PBE functional results in a significant redshift, the ωB97xd in a significant blueshift vs. the experimental data, while B3LYP provides what is usually considered to be a quantitatively accurate peak position. Importantly, both methods fail to capture the effect of the change in conjugation order between Dye 1 and Dye 2.

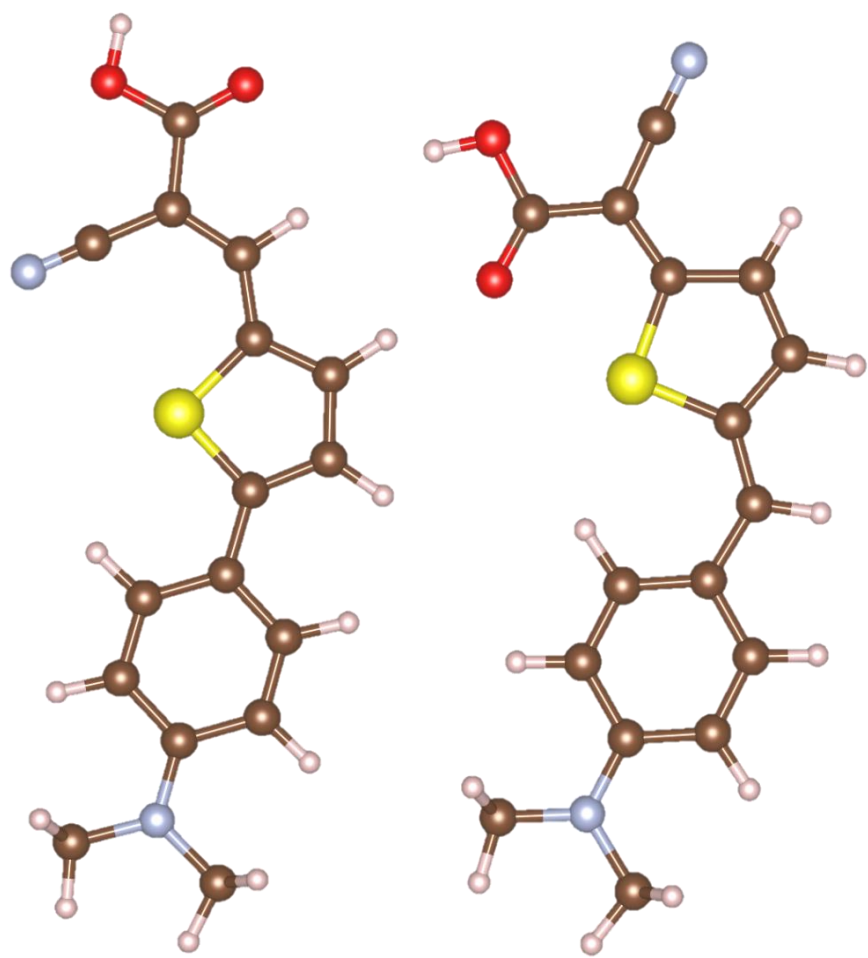

**FIG. 1**.The structures of 2-cyano-3-[5'-(4''-(N,N-dimethylamino) phenyl) thiophen-2'-yl]-acrylic acid (Dye 1, left) and cyano-[5-(4'-(N,N-dimethylamino) benzylidene)-5H-thiophen-2-ylidene]-acetic acid (Dye 2, right). Atom color scheme here and elsewhere: C – brown, H – pink, O – red, N – blue, S – yellow.

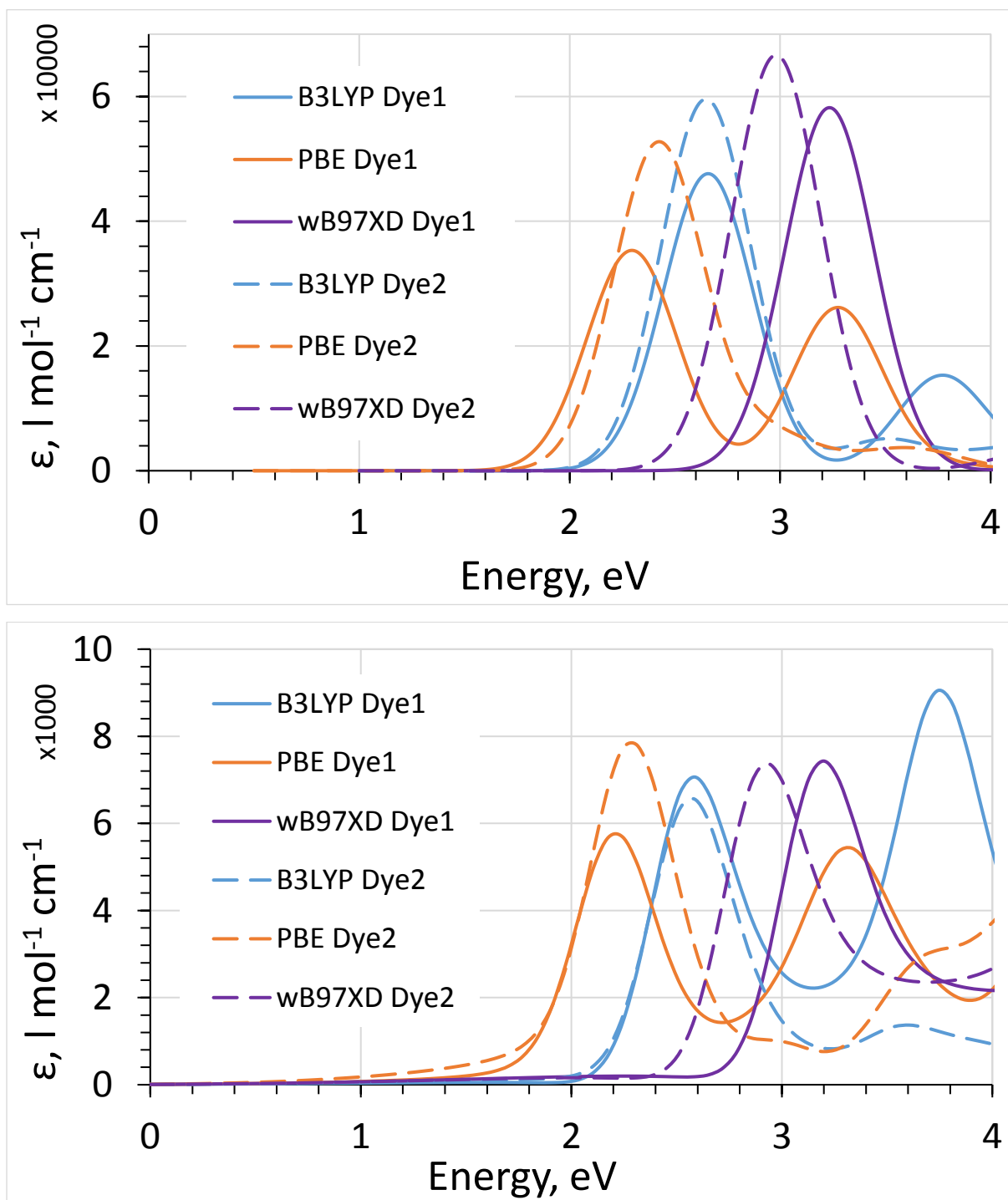

**FIG. 2**. The absorption spectra of Dye 1 and Dye 2 computed with TD-DFT (top panel) and the polarizability-based method (bottom panel).

The intensities of transitions computed with the polarizability-based method are much lower than those computed with TD-DFT. This could be partially related to residual non-Gaussian broadening which is present after the application of Eqs. (5-7) and would lead to lower peak amplitude. However, the values of molar absorptivity on the order of 50,000 L mol$^{-1}$ cm$^{-1}$ following from TD-DFT are typically observed in large rather than small organic dyes. The experimental estimates of molar absorptivity for these molecules in acetonitrile were reported in Ref. [44] and were about 18,000 and 30,000 L mol$^{-1}$ cm$^{-1}$ for Dye 1 and Dye 2, respectively, however, with very wide confidence bands.[44] Overall, there does not appear to be a strong argument in favor of one or the other method as far as intensities are concerned. We conclude that the overall performance of the methods is similar for small molecules.

**Table 1**. Absorption peak maxima in the format "nm (eV)" computed for the small dyes with different methods.

|  | method | Experiment[a] | PBE | B3LYP | ωB97xd |
|---|---|---|---|---|---|
| Dye 1 | TD-DFT | 464 (2.67) | 550 (2.25) | 480 (2.58) | 393 (3.16) |
|  | polarizability |  | 569 (2.18) | 492 (2.52) | 396 (3.13) |
| Dye 2 | TD-TDF | 525 (2.36) | 525 (2.36) | 480 (2.58) | 423 (2.93) |
|  | polarizability |  | 554 (2.24) | 492 (2.52) | 434 (2.86) |
| Ratio | TD-DFT | 1.13 | 0.95 | 1.0 | 1.08 |
|  | polarizability |  | 0.97 | 1.0 | 1.10 |

[a] from Refs. [23, 24]. The experimental spectrum measured in a low-polarity solvent is comparable to calculations performed in vacuum.

**Table 2**. Visible absorption peak maxima in the format "nm (eV)" computed for thiophene oligomers with different methods (B3LYP functional).

|  | TD-DFT | Polarizability |
|---|---|---|
| Dimer | 290 (4.27) | 320 (3.88) |
| Tetramer | 408 (3.04) | 479 (2.59) |
| Octamer | 537 (2.31) | 701 (1.77) |

*Oligomers*

The structures of the thiophene dimer, tetramer and octamer are shown in **FIG. 3**. The absorption spectra computed with TD-DFT and the polarizability-based method are shown in **FIG. 4**, and the corresponding peak positions are listed in **Table 2**. In these calculations, we use the B3LYP functional only, for reasons of CPU cost as well as based on the above results showing that TD-DFT and the polarizability method behave similarly with respect to the choice of the functional. As expected, increasing the length chain leads to a significant redshift. It is well known that with TD-DFT (using GGA or simple hybrid functionals such as B3LYP used here) this redshift is overestimated vs reality. While there are no experimental absorption spectra which would be directly comparable to the computed spectra of oligomers of controlled length and in vacuum, the comparison to experimental spectra is not necessary to judge on the relative performance of the two methods: indeed, it is clear from **FIG. 4** that the redshift is more significant with the polarizability-based method than with TD-DFT. Therefore, the polarizability-based approach underperforms TD-DFT when the extent of conjugation increases. Similar to the case of small molecules above, the intensities are lower several-fold compared to TD-DFT.

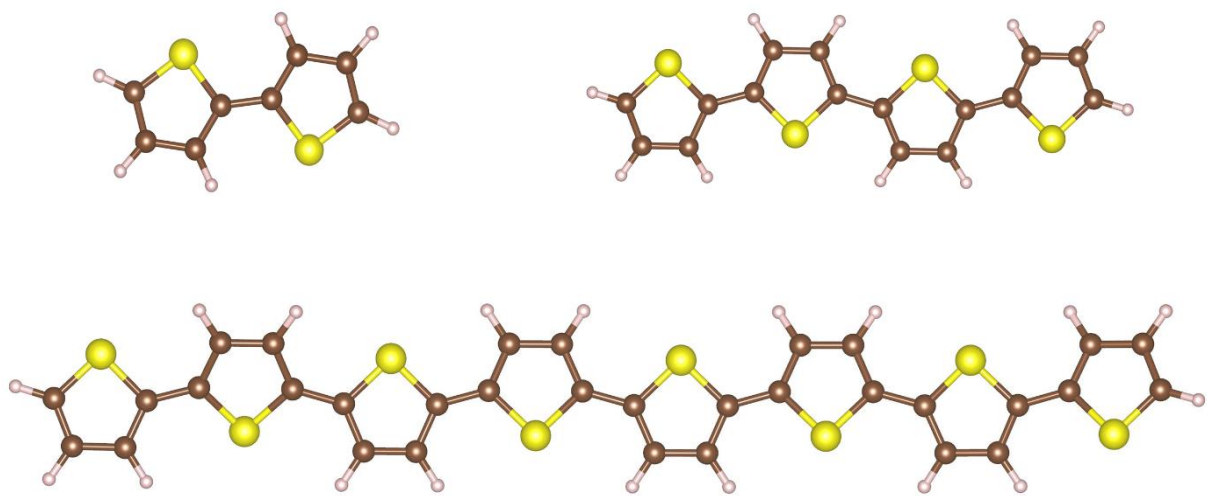

**FIG. 3**. Structures of thiophene oligomers used here.

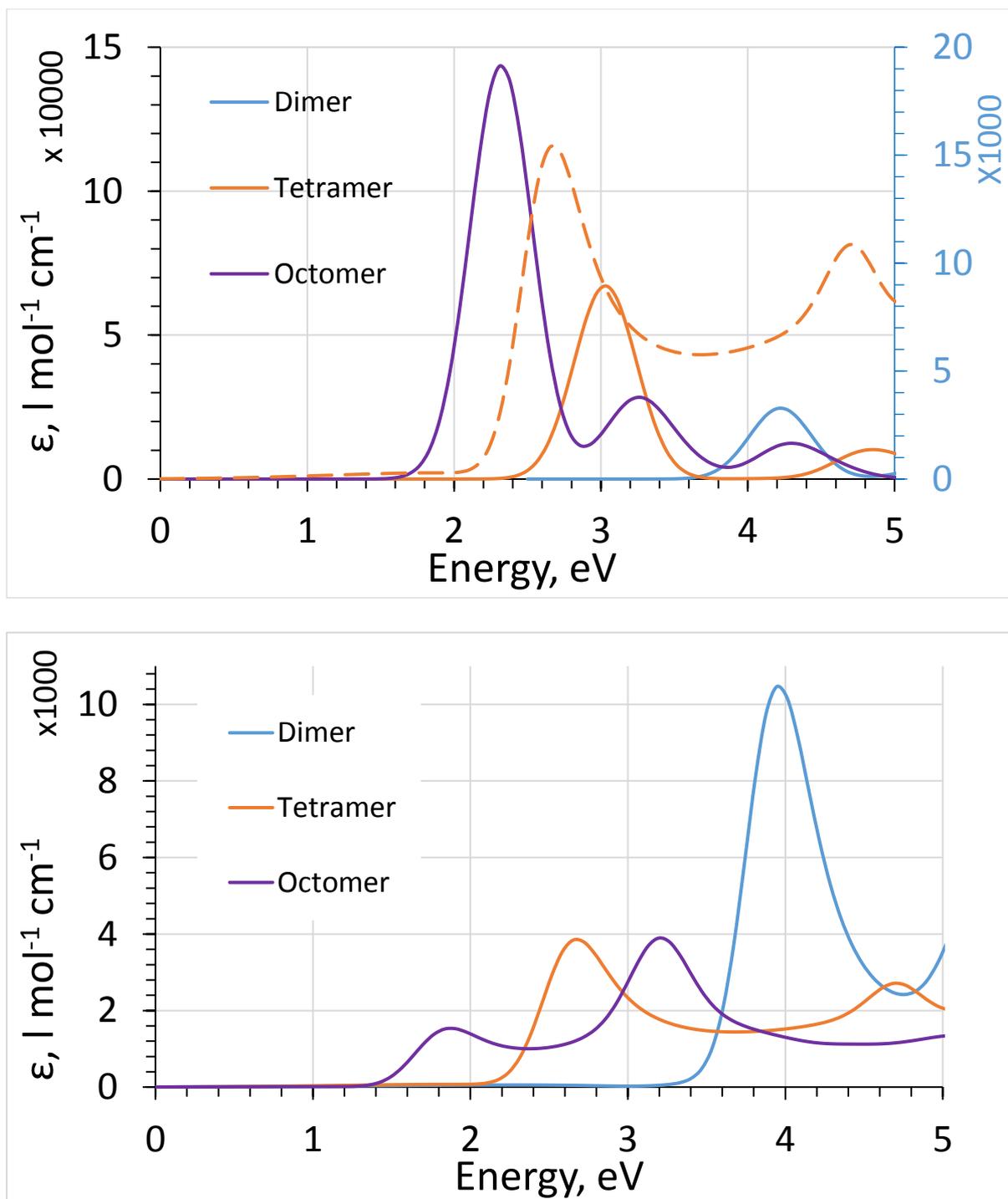

**FIG. 4**. Absorption spectra of thiophene oligomers computed with TD-DFT (top panel) and the polarizability-based method (bottom panel). In the top panel, the dashed curve is the spectrum of the tetramer from the polarizability-based calculation (same as the corresponding curve in the bottom pane), for comparison; the right ordinate axis is for this curve.

*Molecular aggregates and solids*

The crystal structure of pentacene is shown in **FIG. 5**. In the figure, we also highlighted three types of dimers used in the calculations: the herringbone and the stacked dimers in which one expects significant effects of inter-molecular interactions on the absorption spectrum, and the so-called "long" dimer where such interactions are expected to be minimal.

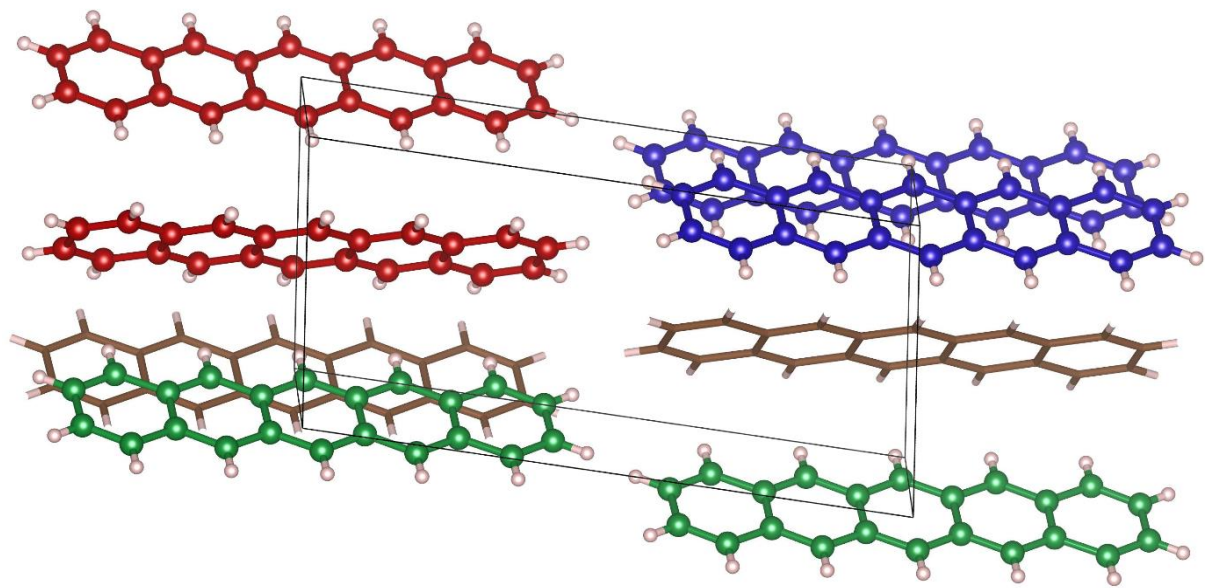

**FIG. 5**. The crystal structure of pentacene. The black line outlines the unit cell. Different dimers used in the calculations are highlighted by color: the red highlights the herringbone configuration, blue the stacked configuration, and green the back-to-back or "long" dimer. The tetramer corresponds to the four molecular units in the left half of the figure.

The absorption spectra of the monomer and the dimers computed using the PBE functional are shown in **FIG. 6**, and using the B3LYP functional in **FIG. 7**. The peak position are listed in **Table 3**. The absorption spectrum of the monomer is dominated by a peak in the visible region (on which we focus here) and a peak in the UV region. The visible peak is dominated by the HOMO to LUMO transition. As expected, the long dimer does not show appreciable aggregation effects on the spectrum, with either method and either functional. The herringbone and the stacked dimers cause very different changes in the spectrum vs. the monomer. We first consider the spectra computed with PBE. With TD-DFT, the stacked dimer causes a modest redshift, a shoulder peak appearing at about 1.2 eV vs 1.7 eV for the peak in the monomer. The herringbone dimer, on the other hand, causes the appearance of a peak downshifted in the excitation energy by about *a half*. This is clearly unrealistic; for example, available experimental spectra of pentacene in solution (which can be used as a proxy for the single-molecule spectrum) and solid state suggest a redshift due to aggregation on the order of 15%.[45]

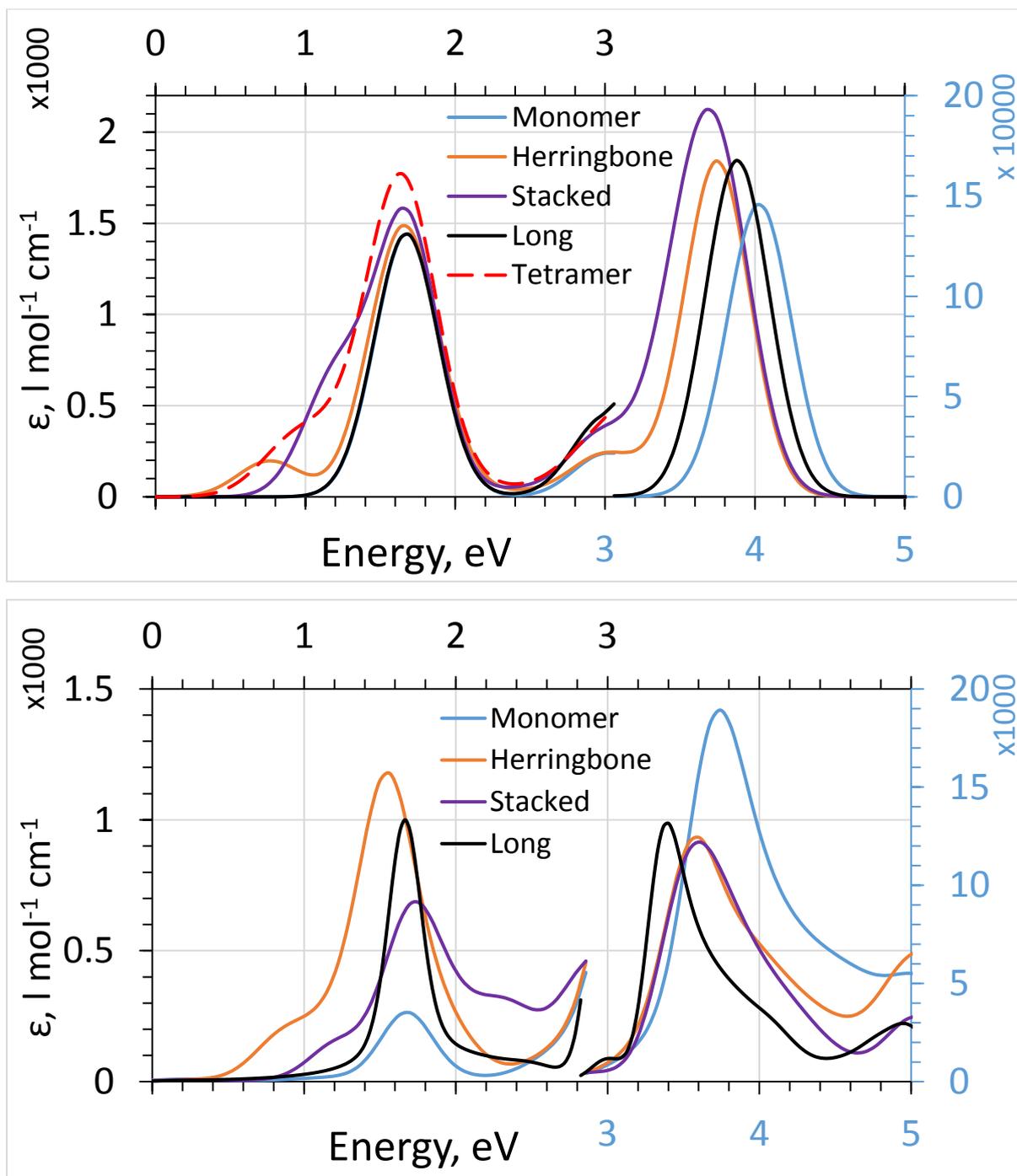

**FIG. 6**. Absorption spectra of pentacene monomer and dimers computed with TD-DFT (top panel) and the polarizability based method (bottom panel) using the PBE functional. The left abscissa axis refers to absorption energies range of the visible peak, and the right axis to that of the UV peak, except for the herringbone and parallel dimers computed with TD-DFT (top panel).

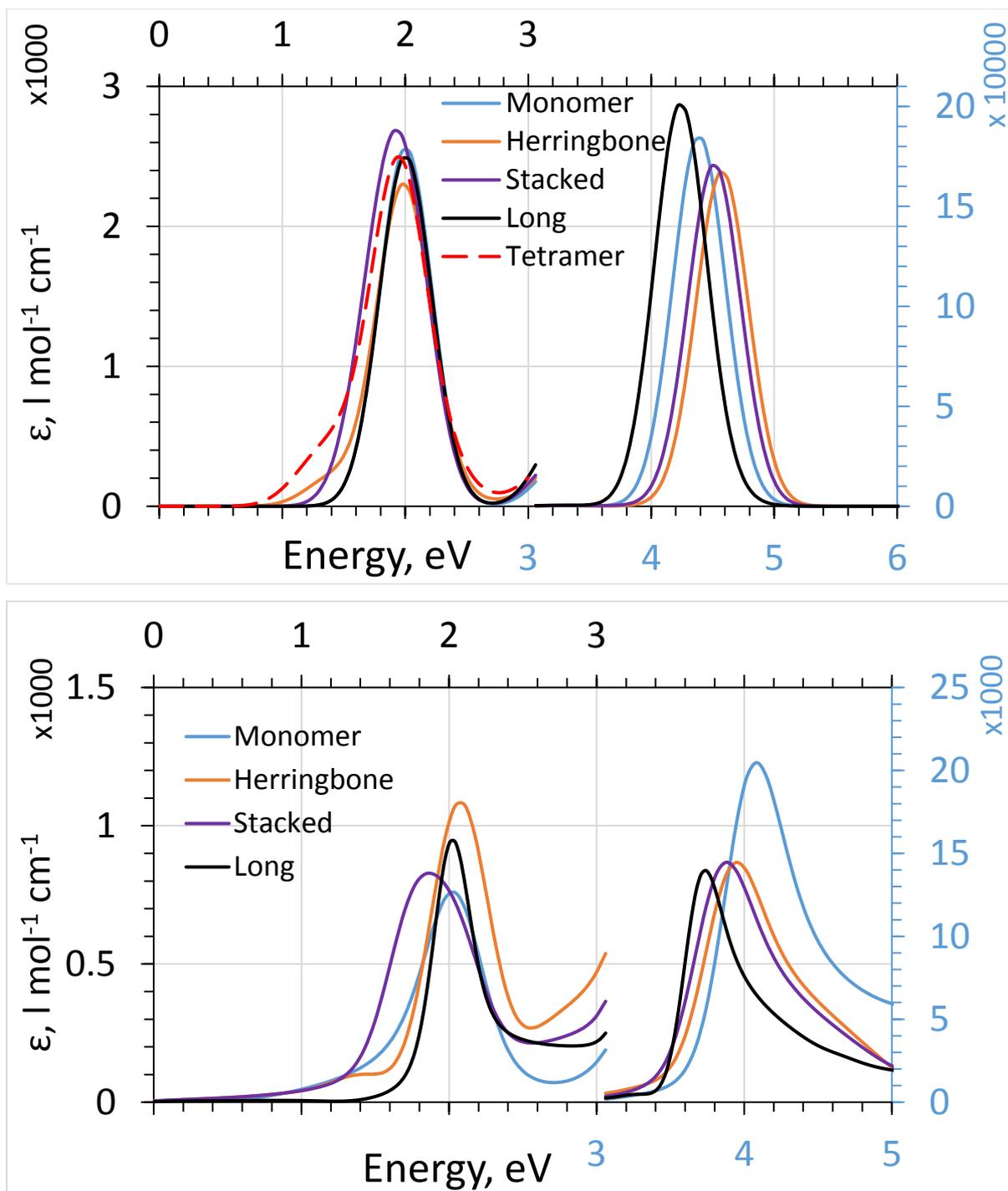

FIG. 7. Absorption spectra of pentacene monomer and dimers computed with TD-TFT (top paned) and the polarizability based method (bottom panel) using the B3LYP functional. The left abscissa axis refers to absorption energies range of the visible peak, and the right axis to that of the UV peak.

**Table 3**. Absorption peak maxima in the format "nm (eV)" computed for pentacene clusters with different methods.

|  | TD-DFT | | Polarizability | | Dipole approx. |
|---|---|---|---|---|---|
|  | B3LYP | PBE | B3LYP | PBE | PBE |
| Monomer | 608 (2.04) | 729 (1.70) | 608 (2.04) | 729 (1.70) | 1097 (1.13) |
|  | 281 (4.42) | 309 (4.01) | 309 (4.01) | 338 (3.67) | 430 (2.88) |
| Herringbone dimer | 886 (1.40) | 1512 (0.82) | 867 (1.43) | 1512 (0.82) | 1097 (1.13) |
|  | 629 (1.97) | 729 (1.7) | 588 (2.11) | 790 (1.57) | 430 (2.88) |
|  | 272 (4.56) | 332 (3.74) | 314 (3.95) | 343 (3.61) |  |
| Stacked dimer | 701 (1.77) | 1033 (1.20) | 701 (1.77) | 1016 (1.22) | 1216 (1.02) |
|  | 629 (1.97) | 729 (1.7) | 653 (1.904) | 729 (1.7) | 1097 (1.13) |
|  | 272 (4.56) | 338 (3.67) | 320 (3.88) | 343 (3.61) | 1000 (1.24) |
|  |  |  |  |  | 452 (2.74) |
|  |  |  |  |  | 443 (2.80) |
|  |  |  |  |  | 435 (2.85) |
|  |  |  |  |  | 429 (2.89) |
| Long dimer | 608 (2.04) | 729 (1.7) | 620 (2.00) | 743 (1.67) |  |
|  | 294 (4.22) | 320 (3.88) | 338 (3.67) | 372 (3.33) |  |
| Tetramer | 1007 (1.23) | 1504 (0.82) |  |  | 1722 (0.72) |
|  | 938 (1.32) | 1485 (0.83) |  |  | 1459 (0.85) |
|  | 825 (1.50) | 1242 (0.99) |  |  | 1278 (0.97) |
|  | 720 (1.72) | 1152 (1.07) |  |  | 1097 (1.13) |
|  | 716 (1.73) | 817 (1.52) |  |  | 976 (1.27) |
|  | 686 (1.80) | 803 (1.54) |  |  | 446 (2.78) |
|  | 650 (1.91) | 746 (1.66) |  |  | 432 (2.87) |

|  |  |  |
|---|---|---|
|  | 632 (1.96) | 697 (1.78) |
|  | 608 (2.04) |  |
|  | 582 (2.13) |  |
|  | 536 (2.31) |  |

| | |
|---|---|
| Crystal | 6526 (0.19) |
|  | 4133 (0.30) |
|  | 2883 (0.43) |
|  | 1771 (0.70) |
|  | 1670 (0.79) |
|  | 1348 (0.92) |
|  | 1170 (1.06) |
|  | 947 (1.31) |
|  | 435 (2.85) |

This behavior of the calculation can be understood by considering orbital localizations, which are shown in **FIG. 8** for the two types of dimers. The orbital localizations in these dimers are qualitatively similar with PBE and B3LYP. The frontier orbitals of the herringbone dimer are localized on individual molecular units and are similar to the HOMO and LUMO of monomers (not shown). This gives rise to charge transfer transitions between the HOMO of one molecule to the LUMO of another, which are strongly and artificially redshifted. The orbitals of the stacked dimer are delocalized over both molecules and the resulting redshift is more in line with the band gap contraction upon dimerization (from 1.97 eV to 1.90 eV). Contrary to the case of the herringbone and stacked dimer, the orbital distributions for the long dimer are different with PBE and B3LYP, they are shown in **FIG. 9**. Specifically in the case of PBE, there is strong localization on individual molecules, which has the potential to cause a strong and artificial redshift via charge transfer transitions. However, in the long dimer, the overlap of orbitals centered on different molecules is too small to appreciably affect the spectrum shown in **FIG. 6**, although the charge transfer transition (at 1.19 eV) is formally present with an oscillator strength of 0.

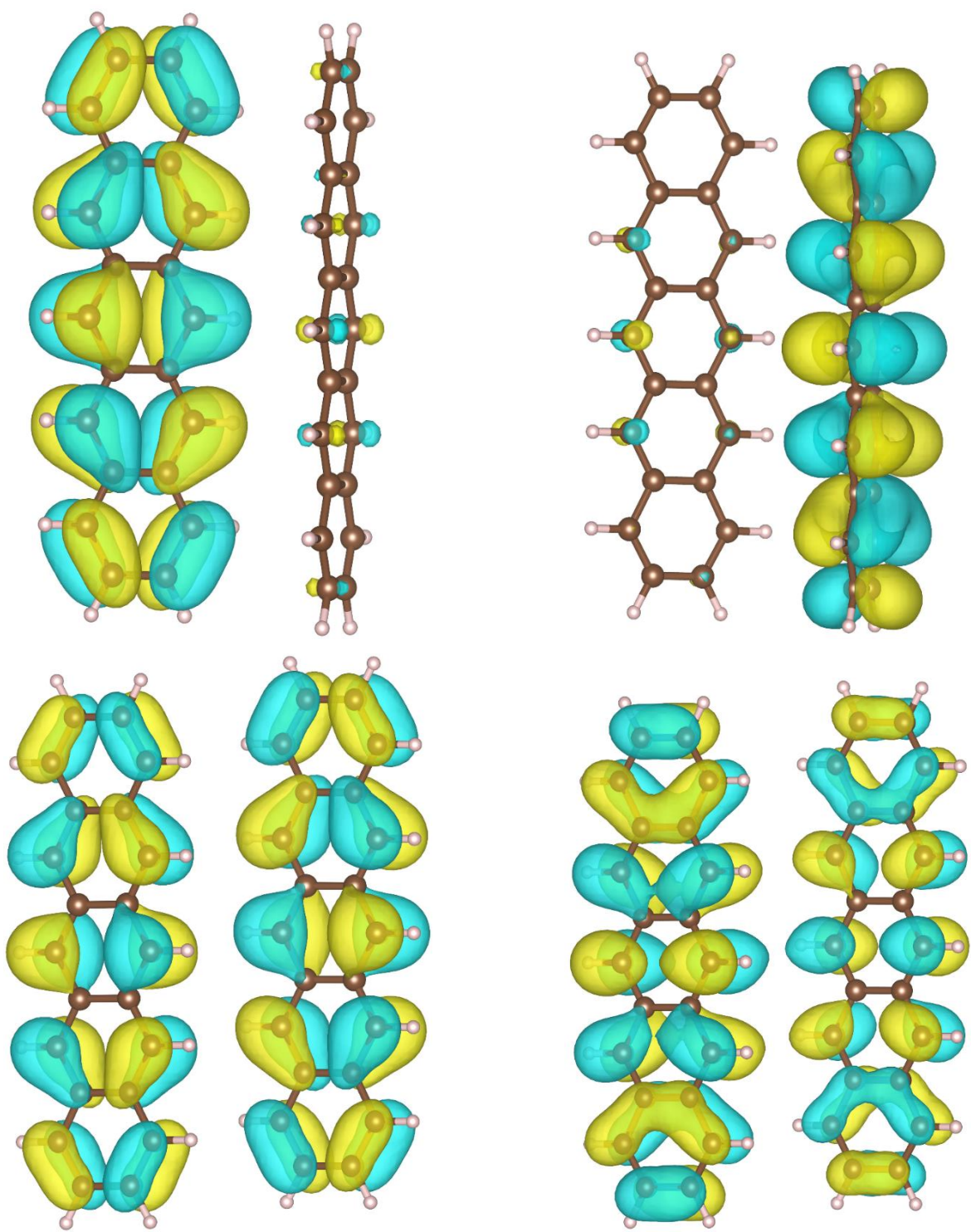

**FIG. 8**. Top panels: HOMO (left) and LUMO (right) of the herringbone dimer. The HOMO-1 and LUMO+1 orbitals are, respectively, HOMO- and LUMO-like and are located on the other monomer. Bottom panels: HOMO (left) and LUMO (right) of the stacked dimer. HOMO-1 and LUMO+1 are similarly delocalized over both units. The orbital localizations are similar with PBE and B3LYP.

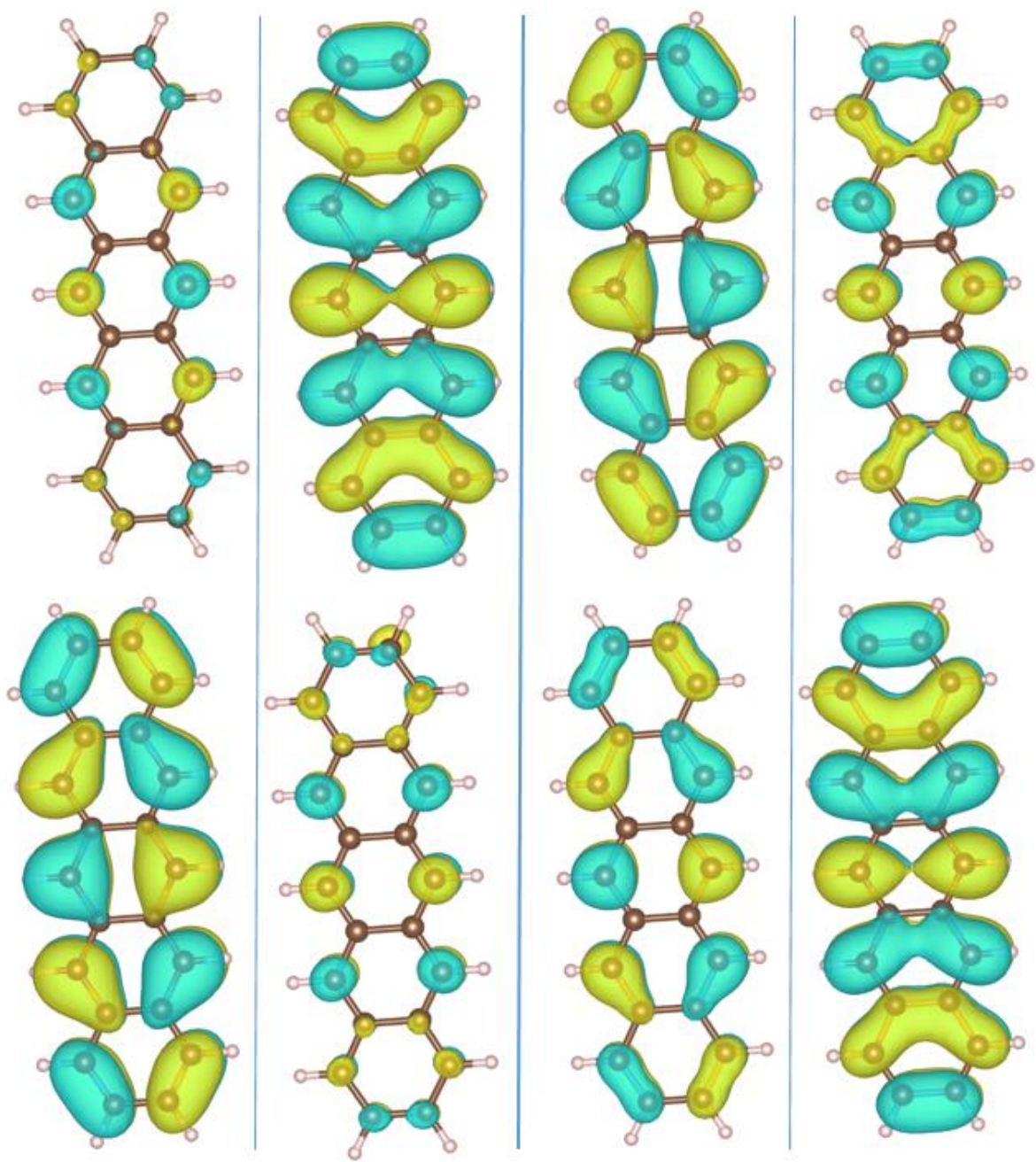

**FIG. 9**. HOMO (left in each pair) and LUMO (right in each pair) orbitals of the long pentacene dimer computed with PBE (left pair of dimers) and B3LYP (right pair of dimers).

The spectra with the polarizability-based method are similar to those computed with TD-DFT. The position of the visible peak of the monomer is practically the same. The intensities are lower with the polarizability based method as in the previous cases. In the herringbone dimer, there is a strongly redshifted peak whose position is the same (0.82 eV) as with TD-DFT. The relative redshift is however smaller as the main visible peak is at 1.57 eV vs 1.70 eV with TD-DFT. While with TD-DFT, the main visible peak of both the stacked and the herringbone dimer is practically the same as in the monomer, with the polarizability based method, the visible peak in the stacked dimer is slightly blueshifted while that of the herringbone dimer is slightly redshifted vs the monomer. This however may have to do with effects of non-Gaussian broadening introduced by Eqs. 6-7.

The performance of the TD-DFT method with the hybrid functional (**FIG. 7**) is as expected in that the charge transfer transition shoulder in the herringbone dimer is less red-shifted, with a shoulder peak at 1.40 eV vs the main peak at 2.04 eV. The stacked dimer has a slightly redshifted transition at 1.77 eV which does not appear as a separate peak under broadening, so that the overall absorption peak appears slightly redshifted. These spectral features are also similar with the polarizability-based method, which, however results in a slightly stronger redshift for the stacked dimer. In the herringbone dimer, the polarizability methods results in a shoulder peak (charge transfer transition) at 1.43 eV vs 2.11 eV for the main peak. The polarizability-based methods, here too, results in lower intensities. The performance is by and large similar with the two methods with the B3LYP functional.

We also performed TD-DFT calculations of the tetramer (we do not perform the tetramer calculation with the polarizability based method due to a significant CPU cost). The tetramer contained the herringbone and the stack dimer, see **FIG. 5**. With PBE, the peaks at 1.66 eV (major) and 0.82 (charge-transfer) eV are similar to those with the herringbone dimer. The redshifted shoulder peak which was observed with the stacked dimer is merged with the main peak due to a larger number of transitions in the larger cluster. With B3LYP, the main peak at 2.04 eV and the redshifted shoulder at 1.23 eV are also similar to the herringbone dimer, with the contributions from the stacked dimer have the visual effect of broadening of the main peak. Orbital localizations for orbitals most responsible for visible transitions are shown in the Supporting Information, FIGS. S1-S4. These localizations confirm the charge-transfer character of the redshifted peak which is mostly due to a HOMO-to-LUMO transition. Orbital localization does not necessarily follow that observed in dimers, for example, some orbitals have appreciable amplitude over three molecular units.

We also performed the absorption spectrum calculations of pentacene monomers, stacked and herringbone dimers, the tetramer, and crystalline pentacene in a fully periodic model, using the PBE functional and the dipole approximation. The results are shown in **FIG. 10**. The peak positions of underlying transitions are listed in **Table 3**. The monomer peak is at about 1.2 eV and is therefore much redshifted vs TD-DFT (as well as vs the polarizability-based) calculations with the same functional. The stacked dimer's visible peak is only slightly redshifted vs the monomer spectrum. The visible peak of the herringbone dimer is at the same excitation energy as in the monomer, and there is no red-shifted peak. This is in spite of the fact that the orbital localizations in the dimers are similar to those in Gaussian calculations (see Supporting Information FIG. S5). The peak in the tetramer is further broadened by the presence of a larger number of transitions but does not show a strongly redshifted charge transfer transition peak as in TD-DFT with PBE. The visible peak of the crystal is strongly redshifted. The peak in **FIG. 10** is made of several components listed in **Table 3**. The redshift is unrealistic based on the available experimental evidence.[45, 46] In the periodic crystal, frontier orbitals are delocalized over all molecular units (Supplementary Information FIG. S6) i.e. orbital localization does not correspond to photoexcited electron distribution in this excitonic material; although in itself this need not imply DFT or TD-DFT failure (see discussion in Ref. [33]), this will impact their accuracy through errors due to DFT approximations. The absence of a significant spectral shift for dimers and an enormous, unrealistic redshift for the crystal also implies absence of uniform convergence of the spectrum with the increasing size of the molecular aggregate, which is likely a method failure.

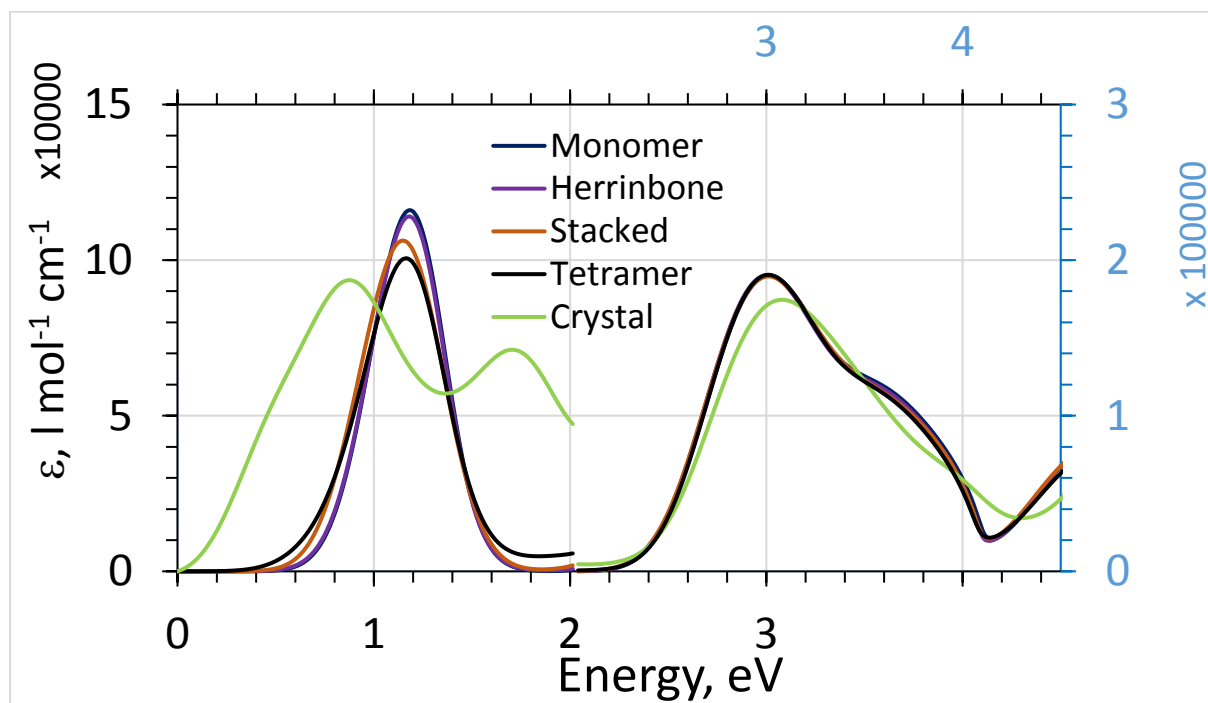

**FIG. 10**. Absorption spectra of pentacene monomer and dimers computed with the dipole approximation using the PBE functional. The left abscissa axis refers to absorption energies range of the visible peak, and the right axis to that of the UV peak.

**Discussion and conclusions**

We have examined the performance of the recently proposed approach to compute absorption spectra of organic materials based on real frequency-dependent polarizability. Contrary to our previous work,[33] we here explicitly focused on molecules where visible peaks are dominated by transitions between HOMO and LUMO. We considered several major classes of organic materials which are known to pose difficulties when computing optical spectra: small molecules with changed conjugated order, oligomers of different lengths (oligothiophene), and molecular aggregates of different size (pentacene). The performance of the polarizability based method for these materials was compared to the commonly used TD-DFT and dipole approximations (for pentacene) as well as to available experimental data, where applicable.

The approach based on real polarizability produces spectra in good agreement with TD-DFT for small molecules. The (artificial) redshift for longer oligomers is more significant with the polarizability-based method than with TD-DFT. For pentacene clusters, TD-DFT produces reasonable spectra with a hybrid functional, but a significant redshift is introduced with a GGA functional due to the presence of charge transfer transitions. This problem is only slightly attenuated with the polarizability-based method. The dipole approximation results in spectra much redshifted vs both TD-DFT and the polarizability-based method and in a different trend with cluster size. The spectrum of solid pentacene computed with the dipole approximation was strongly and unrealistically redshifted vs monomer, dimers, and tetramer; there appeared to be no convergence trend vs cluster size.

The polarizability-based method is well parallelizable, however, used with currently available methods and software tools it also has some serious disadvantages. One difficulty in applying this method is the convergence of the Kramers-Kronig integral. The integrand must span the energy range of dozens of eV to achieve convergence. At high energies, the polarizability calculations are not convergent. We dealt with this issue by computing $\alpha(\omega)$ up to 10 eV and then tailing it to zero up to 20 eV. This still results in some residual non-Gaussian broadening of the spectrum which on one hand has the effect of lowering peak maxima and on the other hand of slightly shifting peak positions when several peaks are in proximity. The current implementation of the method does depend on Kohn-Sham orbitals albeit in a different way than in the TD-DFT and dipole approximations. The convergence at high excitation energy, as well as dependence on the orbitals is something that should be tackled when developing new approximations for $\alpha(\omega)$. We hope that the use of $\alpha(\omega)$ to compute optical absorption spectra of molecules and molecular aggregates that we presented here as well as in Ref. [33] will encourage such research.

# SUPPLEMENETARY INFORMATION
# Comparison of three methods to compute optical absorption spectra of organic molecules and solids

Ang Siong Tuan, Amrita Pal, Sergei Manzhos[1]

Department of Mechanical Engineering, National University of Singapore, Block EA #07-08, 9 Engineering Drive 1, Singapore 117576

---

[1] Author to whom correspondence should be addressed. Tel: +65 6516 4605; fax: +65 6779 1459; E-mail: mpemanzh@nus.edu.sg

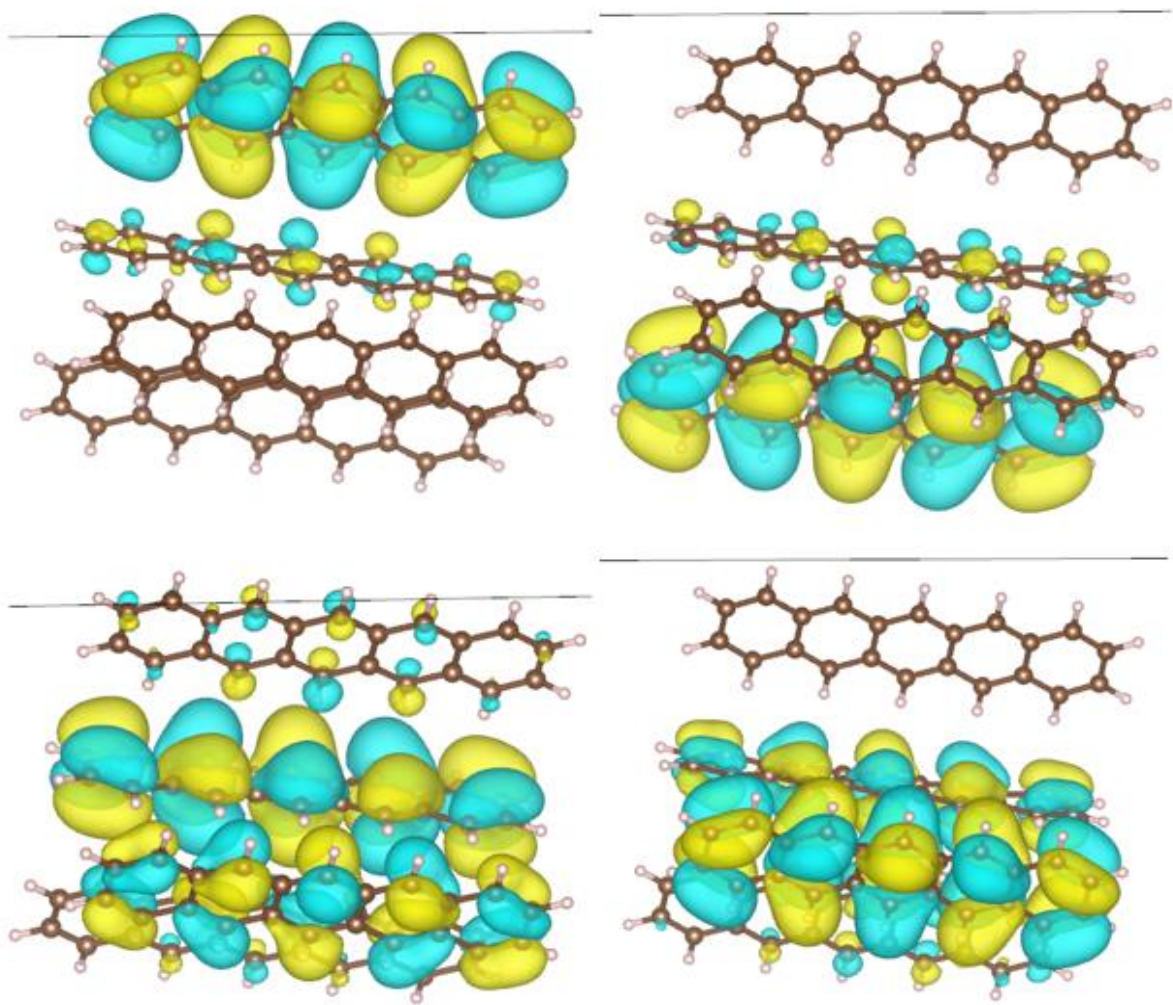

FIG. S1. Left to right and top to bottom: HOMO, HOMO-1, HOMO-2, HOMO-3 orbitals of the pentacene tetramer computed with the PBE functional.

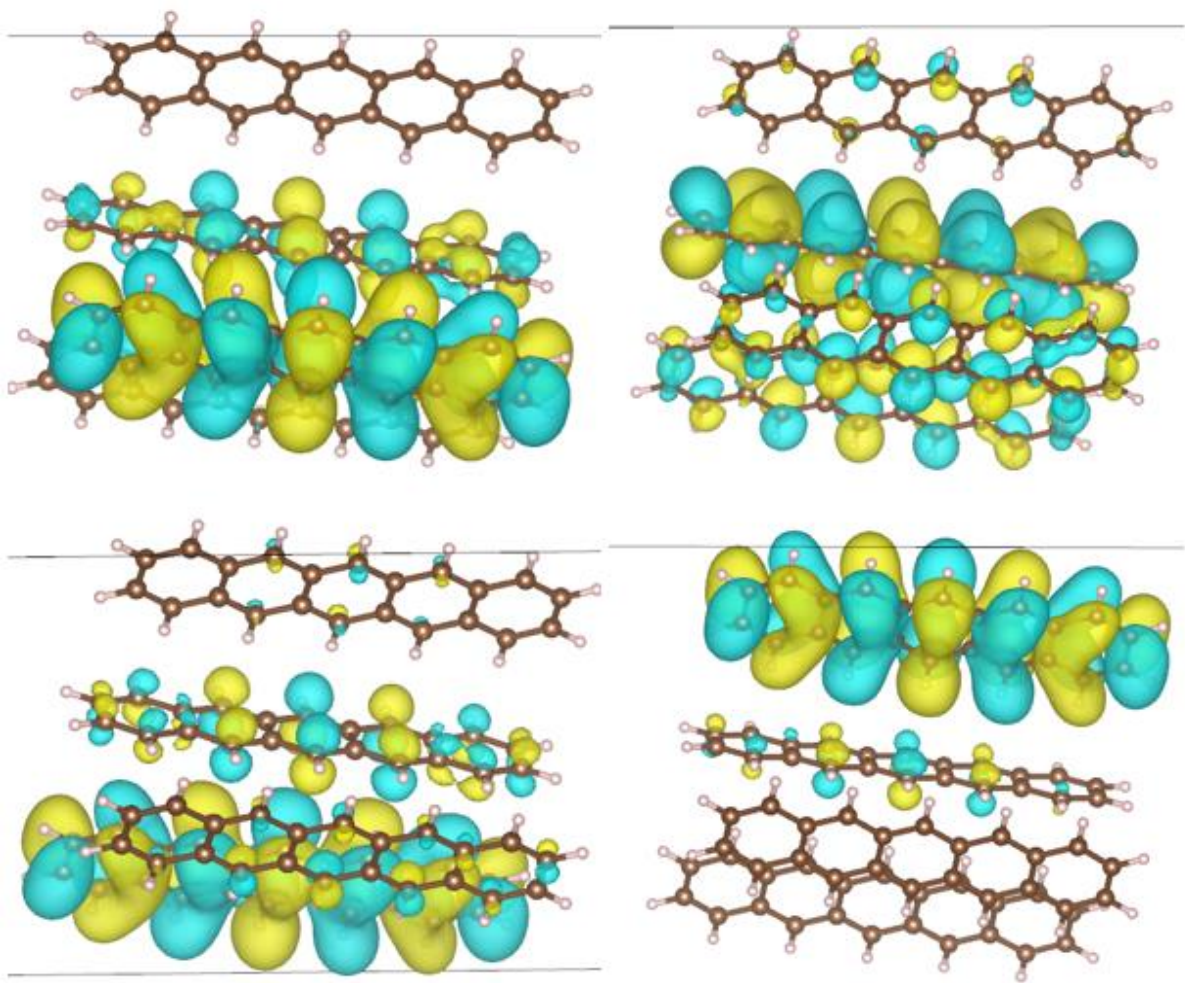

FIG. S2. Left to right and top to bottom: LUMO, LUMO+1, LUMO+2, LUMO+3 orbitals of the pentacene tetramer computed with the PBE functional.

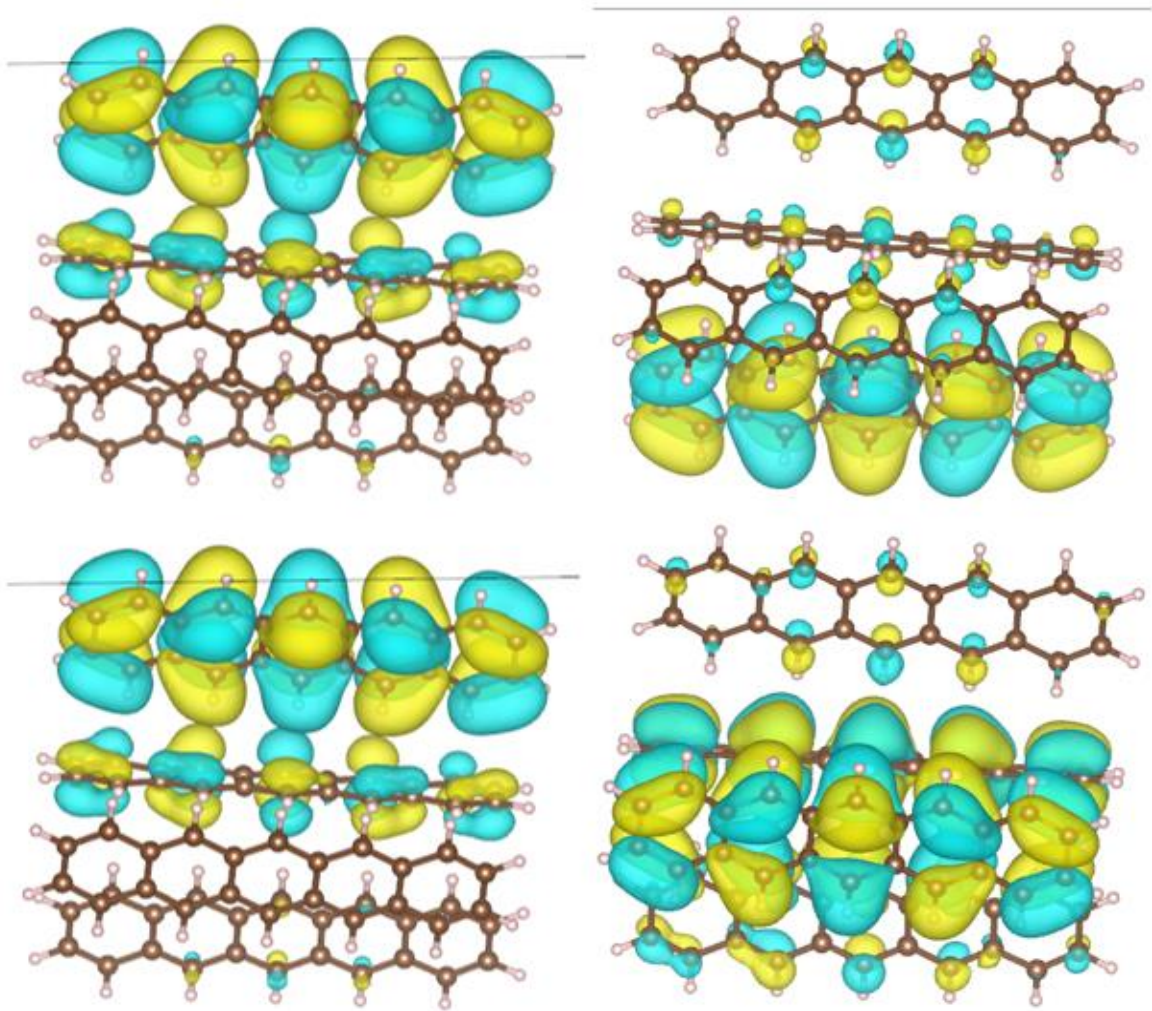

FIG. S3. Left to right and top to bottom: HOMO, HOMO-1, HOMO-2, HOMO-3 orbitals of the pentacene tetramer computed with the B3LYP functional.

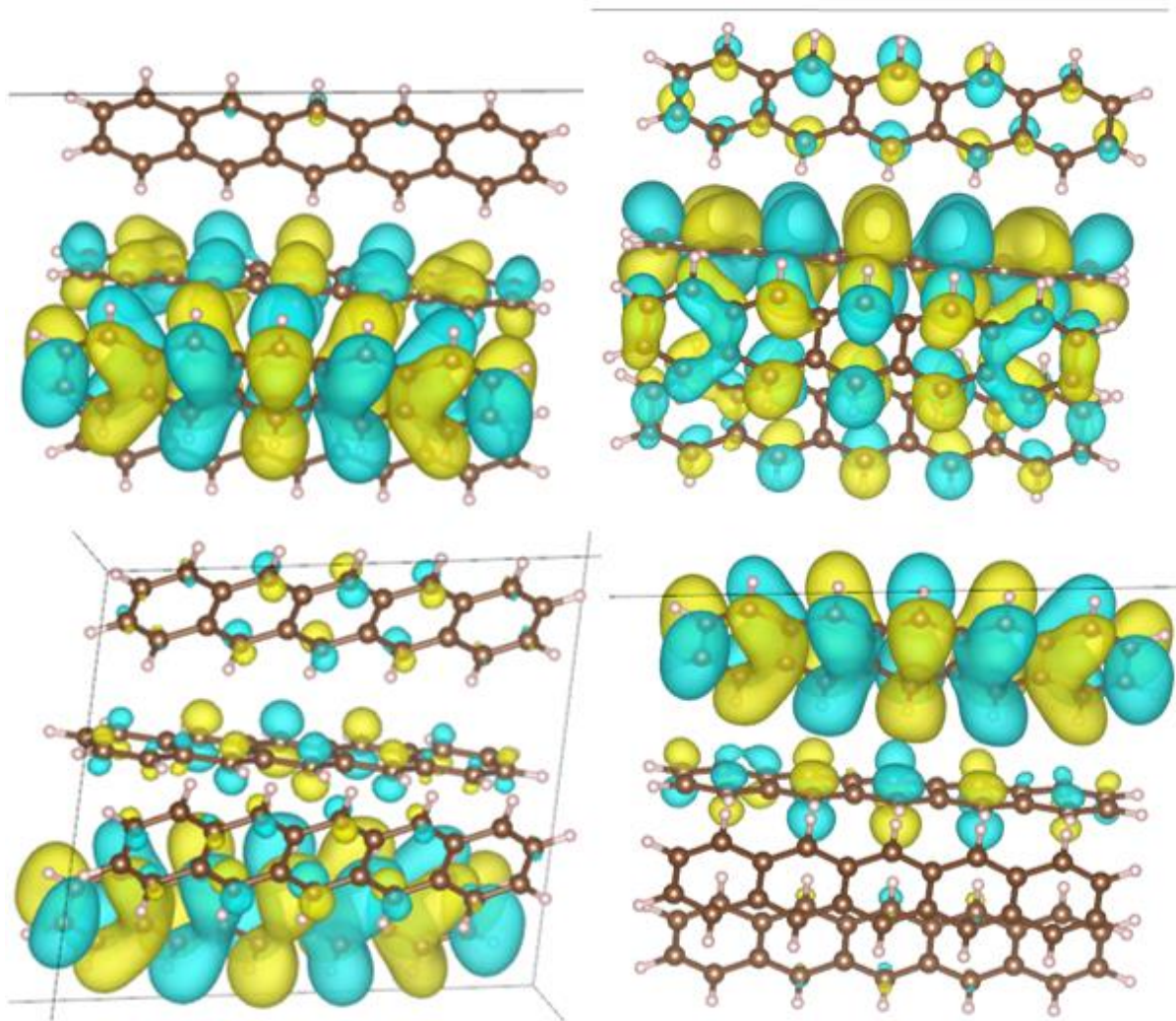

FIG. S4. Left to right and top to bottom: LUMO, LUMO+1, LUMO+2, LUMO+3 orbitals of the pentacene tetramer computed with the B3LYP functional.

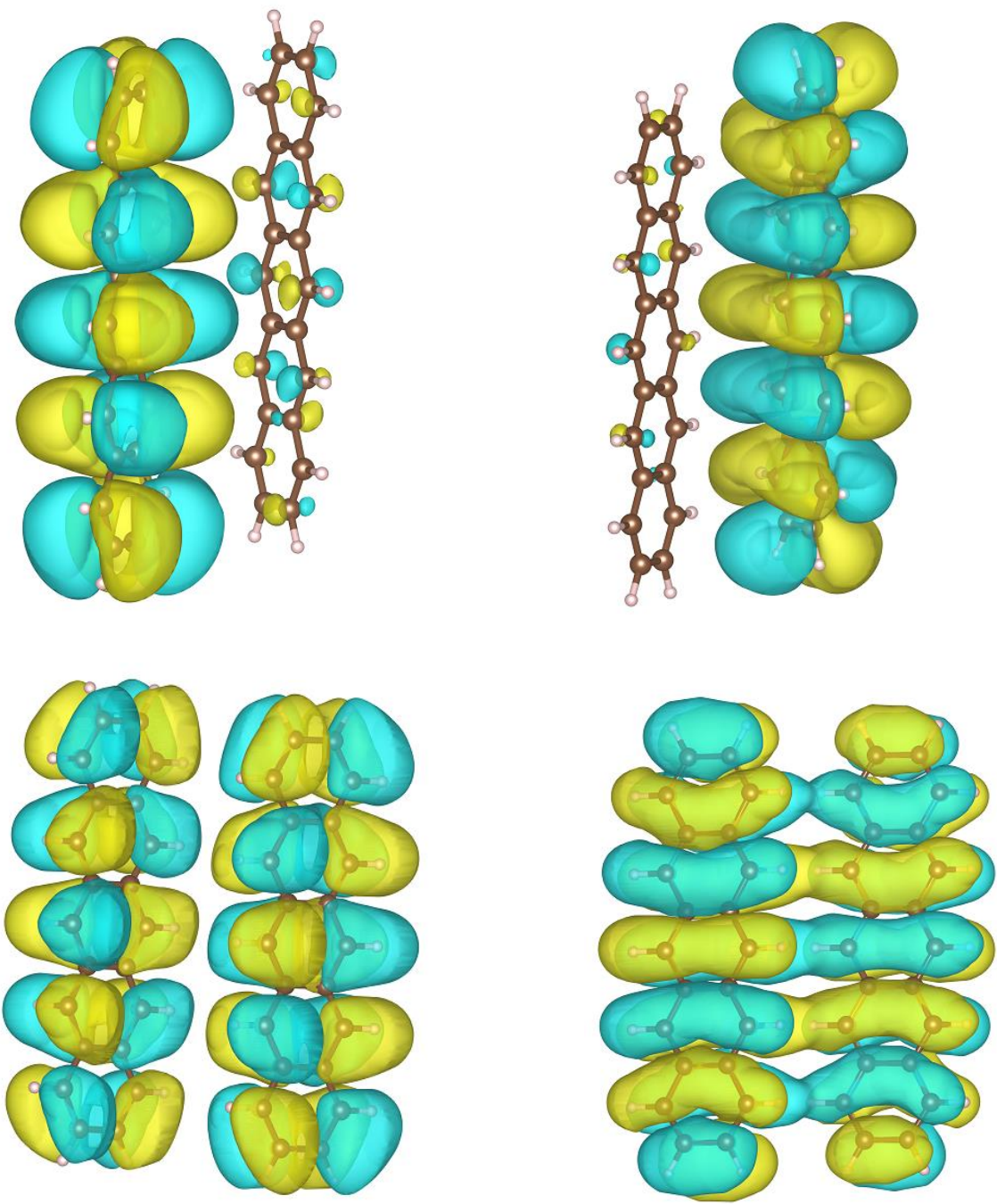

FIG. S5. Top panels: HOMO (left) and LUMO (right) of the herringbone dimer. The HOMO-1 and LUMO+1 orbitals are, respectively, HOMO- and LUMO-like and are located on the other monomer. Bottom panels: HOMO (left) and LUMO (right) of the stacked dimer. HOMO-1 and LUMO+1 are similarly delocalized over both units. The orbital are computed with PBE in SIESTA.

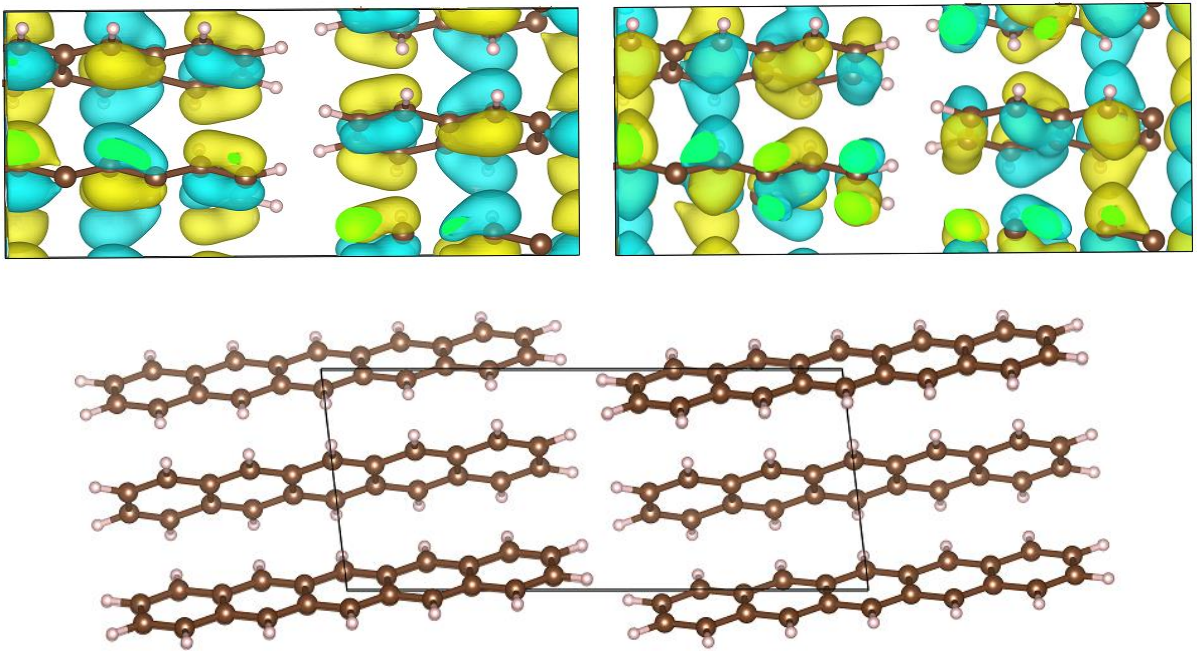

FIG. S6. HOMO (top left) and LUMO (top right) at the Γ point of the pentacene crystal computed with the PBE functional in SIESTA. A rectilinear cut of unit cell is shown. For comparison, the cell with molecules completed beyond the unit cell is shown at the bottom.